\documentclass[journal,draftcls,onecolumn,11pt]{article}
  \usepackage{epsfig}
\begin{document}
\title{On the solutions of the minimum energy problem in one dimensional sensor networks}
\author{
Z. Lipi{\'n}ski  \\
        Institute of Mathematics and Informatics \\
        Opole University, Poland
}
\maketitle
\begin{abstract}
We discuss solutions of the minimum energy problem in one dimensional wireless sensor networks
for the data transmission cost function $E(x_i,x_j) = d(x_i,x_j)^a + \lambda \;d(x_i,x_j)^b$
with any exponent $a,b\in R$ and $\lambda \geq 0$,
where $d(x_i,x_j)$ is a distance between transmitter and receiver.
We define the minimum energy problem
in terms of sensors signal power, transmission time and capacities of a transmission channels.
We prove, that for the point-to-point data transmission method utilized by the sensors in the physical layer,
when the transmitter adjust the power of its radio signal to the distance to the receiver,
the solutions of the minimum energy problem written in terms of data transmission cost function and
in terms of the sensors signal power coincide.
\end{abstract}
\section{Introduction}
Characteristic feature of sensor networks is that these consist of small electronic
devices with limited power and computational resources.
Typical activity of sensor network nodes is collection of sensed data,
performing simple computational tasks and transmission of the resulting data to a fixed set of data collectors.
The sensors utilize most of their energy in the process of data transmission,
this energy grows with the size of the network and the amount of data transmitted over the network.
Generally  in sensor networks there are two models of energy consumption.
The objective of the first model is to maximize the functional lifetime of the network, \cite{Chang, Giridhar}.
For this type of problems the data transmission in the network is modeled in such a way,
that the energy consumed by each sensor is minimal.
To extend the network lifetime the sensors must share their resources and cooperate in the process of data transmission.
For typical solutions of such problems the consumed energy is evenly distributed over all nodes of the network.
The second type of problems is to optimize the energy consumed by the whole network, \cite{Acharya, Halpern, Rodoplu, Wu, Srinivas}.
Such problems arise when the network nodes are powered by the central source of energy
or the node batteries can be recharged and the total energy consumed by the network is to be minimized.
To solve this type of problems it is enough
to find the optimal energy consumption model of each sensor and summed up their energies.
In this paper we discuss solutions of the minimum energy problem in one dimensional wireless sensor network $S_N$
when each sensor generates the amount $Q_i$ of data and sends it, possible via other sensors, to the data collector.
We assume, that the network $S_N$ is build of $N$ sensors and one data collector.
The sensors are located at the points $x_i>0$ of the line and the data collector at the point $x_0=0$.
We prove, that for any exponent $a\in R$ in the data transmission cost function
\begin{equation} \label{DataTransCostEnergyMat-SN}
E(x_i,x_j)=|x_i - x_j|^a,
\end{equation}
where $|x_i-x_j|$ is a distance between transmitter and receiver,
when in the interval $(x_0, \frac{1}{2}x_N)$ there are $N'$ sensors,
there are $(N'+1)$ solutions of the problem.
We show, how to determine the optimal solutions of the minimum energy problem for
the data transmission cost function which is sum of two factors
\begin{equation} \label{MixedDataTransCostEnMat-SN}
E(x_i,x_j)=|x_i - x_j|^a + \lambda  |x_i - x_j|^b,
\end{equation}
where $a, b\in R$ and $\lambda \geq 0$.
We also discuss the solutions of the minimum energy problem in wireless sensor networks
when the energy utilized by the network is expressed in terms of node signal power, transmission time
and there are constraint on the transmission channels due to presence of the noise and interference, \cite{Baccelli}.

We represent a sensor network $S_N$ as a directed, weighted graph $G_N=\{S_N, V, E \}$ in which
$S_N$ is a set of graph nodes, $V$ is the set of edges and $E$ set of weights.
Each directed edge $T_{i,j}\in V$ defines a communication link between i-th and j-th node of the network.
To each edge $T_{i,j}$ we assign a weight $E(x_i,x_j)\equiv E_{i,j}$, which is
the cost of transmission of one unit of data between
$i$-th and $j$-th node.
The data flow matrix $q_{i,j}$ defines the amount of data transmitted along the edge  $T_{i,j}$.
By $U^{({\rm out})}_{i} \subseteq S_N$ we denote a set of the network nodes
to which the $i$-th node can send the data, i.e.,
$U^{({\rm out})}_{i}=\{ j\in S_N | \;\exists \;T_{i,j}\in V \}$.
The set $U^{({\rm out})}_{i}$ defines the maximal transmission range of the i-th node.
In the paper we assume, that each sensor of $S_N$ can send the data to any other node of the network
%
$\forall_{i \in S_N } \;\; U^{({\rm out})}_{i}=S_N$.
%
If we assume, that each sensor of the $S_N$ network generates the amount $Q_i$ of data, $i\in[1,N]$, and
the data is sent to the data collector,
then the energy consumed by the $i$-th sensor in the process of data transmission can be written in the form
$E_{i}(q) = \sum_{j=1}^{N} q_{i,j} E_{i,j}$.
For the total energy consumed by the network
$E_{T}(q) = \sum_{i\in S_N}^{} E_{i}(q)$
the minimum energy problem can be defined by the set of following formulas
\begin{equation} \label{MiniNetEnergyProblem}
\left\{   \begin{array}{l}
\min_{q} \; E_{T}(q), \\
 \sum_{i} q_{i,j} = Q_i + \sum_{j} q_{j,i},\\
 E_{i,j}\geq 0, \;q_{i,j}\geq 0, Q_i > 0\; i,j\in [0,N],
\end{array} \right.\end{equation}
where the second formula defines the feasible set of the problem.
It states that the amount of data generated by the $i$-th node $Q_i$  and the amount of data received from other nodes
$\sum_{j} q_{j,i}$ must be equal to the amount of data which the node can send $\sum_{i} q_{i,j}$.

Because the objective function $E_{T}(q)$ of the problem is continuous and linear we can deduce from this
a simple but helpful fact, that any local minimum of $E_{T}(q)$ is a global one and thus it is a solution of (\ref{MiniNetEnergyProblem}).

If we assume, that we search for a solution of (\ref{MiniNetEnergyProblem}) in the integers, i.e. $q_{i,j}\in Z_{+}^{0}$
for $Q_i\in Z_{+}^{}$, then we get the mixed integer linear programming problem.
It is easy to see that such problem is NP-hard.
To find the minimum of $E_{T}(q)$ first we must find an integer matrix $q$
satisfying the feasible set equation, given by the second relation in (\ref{MiniNetEnergyProblem}).
Because this requires solution of the partition problem, \cite{Garey},
we get the reduction of the partition problem to the minimum energy problem (\ref{MiniNetEnergyProblem}) with the
requirements $q_{i,j}\in Z_{+}^{0}$, $Q_i\in Z_{+}^{}$.
\section{Solution of the problem with the monomial cost function}
In this section we solve the minimum energy problem (\ref{MiniNetEnergyProblem})
for the data transmission cost function (\ref{DataTransCostEnergyMat-SN})
with  arbitrary real value of the exponent $a$.
As can be seen, the monomial (\ref{DataTransCostEnergyMat-SN})
for $a\geq 1$ and $x_i\geq   x_j \geq  x_k\geq 0$ satisfies the inequality
\begin{equation} \label{SuperadditiveDataTransCostEnergMatrDistance}
|x_i - x_j|^a + |x_j - x_k|^a \leq |x_i - x_k|^a,
\end{equation}
and it is an example of a super-additive function, \cite{Steele}.
This is because for $x_i - x_j=x$, $x_j - x_k=y$
(\ref{SuperadditiveDataTransCostEnergMatrDistance}) can be written in the form $|x|^a + |y|^a \leq |x+y|^a$.
Solutions of the minimum energy problem in $S_N$ with the cost function
(\ref{DataTransCostEnergyMat-SN}), where $a\geq 1$, can be easily generalized
to any data transmission cost function $E(x_i,x_j)$ which satisfies the inequality
\begin{equation} \label{SuperadditiveDataTransCostEnergMatr}
\forall_{x_i\geq   x_j \geq  x_k\geq 0}\;\; E(x_i,x_j)+ E(x_j,x_k) \leq E(x_i,x_k).
\end{equation}
From (\ref{SuperadditiveDataTransCostEnergMatr}) it follows that
the energy consumed by each sensor is minimal when it sends all of its data to the nearest neighbor in the direction of data collector.
Let us assume, that the data is transmitted between two nodes located at the points $x_i$ and $x_k$
and (\ref{SuperadditiveDataTransCostEnergMatr}) is satisfied, then the cost of transmission $E(x_i,x_k)$
can be reduced by transmitting the data via the $j$-th node located between them, i.e., via the the point $x_j$ for which
the inequality is satisfied $x_i\geq   x_j \geq  x_k\geq 0$.
Because the total energy consumed by the network is a sum of energies consumed by its nodes,
then the solution of the minimum energy problem
(\ref{MiniNetEnergyProblem}) with (\ref{DataTransCostEnergyMat-SN}) and $a\geq 1$
can be described by the transmission graph
$T^{(0)} = \{ T^{(0)}_{i,i-1}\}_{i=1}^{N},$
with the weight of each edge $T^{(0)}_{i,i-1}$ equal to $q^{(0)}_{i,i-1} = \sum_{j=i}^{N} Q_j$.
The graph $T^{(0)}$ defines the next hop data transmission along the shortest path, where
the shortest path means transmission along the distance $d(x_i,x_j)$ between the transmitter and the receiver.

For $a\leq 1$ elements of the data transmission cost function (\ref{DataTransCostEnergyMat-SN})
are the sub-additive functions, i.e., satisfy the inequality $|x|^a + |y|^a \geq |x+y|^a$, \cite{Steele}.
Solutions of the minimum energy problem (\ref{MiniNetEnergyProblem})
with (\ref{DataTransCostEnergyMat-SN}) and $a\leq 1$
can be generalized to the data transmission cost function $E(x_i,x_j)$ which satisfy the inequality
\begin{equation} \label{SubadditiveDataTransCostEnergMatr}
E(x_i,x_j)+ E(x_j,x_k) \geq E(x_i,x_k).
\end{equation}
The optimal behavior of the sensors which minimizes the total network energy $E_{T}(q)$
can be deduced from the inequality (\ref{SubadditiveDataTransCostEnergMatr}),
but it does not uniquely determine the solution of (\ref{MiniNetEnergyProblem}).
To get the unique solution of (\ref{MiniNetEnergyProblem}) we need a concrete form of the data transmission cost function,
for example (\ref{DataTransCostEnergyMat-SN}) or (\ref{MixedDataTransCostEnMat-SN}).
From (\ref{SubadditiveDataTransCostEnergMatr}) it follows that the cost of transmission between two nodes located
at the points $x_i$ and $x_j$ is minimal when the data is transmitted along the longest hops, i.e.,
any transmission via node which lie between $x_i$ and $x_j$ is less optimal.
For the data transmission cost function (\ref{DataTransCostEnergyMat-SN})
and $a\in (-\infty,1]$ one may expect, that the optimal data transmission is given by the graph
$T^{(1)} = \{ T^{(1)}_{i,0} \}_{i=1}^{N}$,
with the weights $q^{(1)}_{i,0} = Q_{i}$.
This is true for sensors which lie in the interval $[x_N,\frac{1}{2}x_N]$ of $S_N$.
When $x_i\in [x_N,\frac{1}{2}x_N]$, then the distance $d(x_i,x_0)$
between the transmitter and the data collector is maximal and
the inequality (\ref{SubadditiveDataTransCostEnergMatr})
for $x_i$, $x_k=0$ and any sensor $j\in S_N$, i.e., not only for $x_j < x_i$ but also
for $x_j> x_i$, is satisfied.
For sensors which lie in the interval $(0,\frac{1}{2}x_N)$
to find the optimal transmission it must be taken into account
two data transmission paths to the data collector.
The directly to the data collector transmission path
$T^{(1)}_{i}=\{ T^{(1)}_{i,0}\}$ 
and the two hops transmission given by the path
$T^{(1')}_{i} = \{ T^{(1')}_{i,N}, T^{(1')}_{N,0} \}$.
Selection of which one depends on the value of the parameter $a \in (-\infty, 1]$ in (\ref{DataTransCostEnergyMat-SN}).
Let us assume, that in the interval $(0,\frac{1}{2}x_N)$ there are $N'$ sensors.
For each sensor $k$ from the interval $(0,\frac{1}{2}x_N)$
we split the network $S_N$ into two sets $V^{(k)}_{1}$ and $V^{(k)}_{2}$.
To the set
$$V^{(k)}_{1} = \{ i \in S_N\; | \; d(x_i,x_0) < d(x_k,x_0) \},\;\; k\in [1,N']$$
belong sensors which lie to the left the $k$-th sensor.
The set  $V^{(k)}_{2}$ is the completion of $V^{(k)}_{1}$, i.e.,
$V^{(k)}_{2}= S_N \setminus V^{(k)}_{1}$.
The sensors from the interval $(0,\frac{1}{2}x_N)$
can be used to classify solutions of the minimum energy problem
for the data transmission cost function (\ref{DataTransCostEnergyMat-SN})
and any value of the exponent $a\in (-\infty, 1]$.
For the $k$-th sensor from the interval $(0,\frac{1}{2}x_N)$,
we must check whether the optimal data transmission path from the k-th sensor to
the data collector is $\{ T^{(k)}_{k,0}\}$ or $\{ T^{(k)}_{k,N}, T^{(k)}_{N,0} \}$.
In other words, we must check the values of the parameter $a$ for which the inequality holds
%
$$E(x_k,x_N) + E(x_N,x_0) \geq E(x_k,x_0), \;\;\; k\in [1,N'].$$
Instead solving these inequalities, we solve the set of equations
%
$$ E(x_k,x_N) + E(x_N,x_0) - E(x_k,x_0) = 0, \;\;k\in [1,N'],$$
which for (\ref{DataTransCostEnergyMat-SN}) have the form
\begin{equation} \label{xkNxN0-xk0}
|x_N - x_k|^a + x_N^a - x_k^a = 0, \;\; a\in (-\infty, 1], \;k\in [1,N'].
\end{equation}
For $N'$ roots $a_{k}$ of (\ref{xkNxN0-xk0}) we can form $N'$ intervals
$ a\in [a_{k+1}, a_k], \;\; k\in [0,N']$,
where $a_{0}=1$  and $a_{N'+1}=-\infty$.
For any $a\in [a_{k+1}, a_k]$ the following set of inequality holds
\begin{equation} \label{InequalitiesForKandK+1} \left\{   \begin{array}{l}
       |x_N - x_k|^a + x_N^a - x_k^a \leq 0,  \\
       |x_N - x_{k+1}|^a + x_N^a - x_{k+1}^a \geq 0, \\
  \end{array} \right.\end{equation}
which means, that for $a\in [a_{k+1}, a_k]$ the nodes $i\in [1,k]$
transmit data along the two hops path $\{ T^{(k+1)}_{i,N},T^{(k+1)}_{N,0} \}$
and the nodes $i\in [k+1,N]$ along the one hop path $\{ T^{(k+1)}_{i,0}\}$.
The above results summarizes the following

{\bf Lemma 1}.
The solutions of the minimum network energy problem
for the data transmission cost matrix $E_{i,j}= |x_i-x_j|^{a}$ and $a\in R$
is given by the data transmission graphs
\begin{equation} \label{SolutionGraph}
\left\{   \begin{array}{ll}
T^{(0)}=\{ T^{(0)}_{i,i-1}\}_{i=1}^{N} & {\rm for}\;\; a\in [1 ,\infty),\\
T^{(1)}=\{ T^{(1)}_{i,0}\}_{i=1}^{N}  & {\rm for}\;\;  a\in [a_{1},1],\\
T^{(k+1)}=\{ T^{(k+1)}_{i,N}, T^{(k+1)}_{i',0} \}_{i=1,i'=k+1}^{k,N} &{\rm for}\;\;  a\in [a_{k+1},a_{k}], k\in [1,N'-1],\\
T^{(N'+1)}=\{ T^{(N'+1)}_{i,N}, T^{(N'+1)}_{i',0} \}_{i=1,i'=N'+1}^{N',N} & {\rm for}\;\; a\in (-\infty,a_{N'}],
 \end{array} \right. \end{equation}
with the weights
\begin{equation} \label{WeightsForSolutionGraph}
\left\{   \begin{array}{l}
 q^{(0)}_{i,i-1} = \sum_{j=i}^{N} Q_j, \;\; i\in [1,N],\\
q^{(1)}_{i,0} = Q_{i}, \;\; i\in [1,N],\\
 q^{(k+1)}_{i,N} = Q_{i}, i\in [1,k], \;\; q^{(k+1)}_{i',0}=Q_{i'}, i'\in [k+1,N-1],\\
 q^{(k+1)}_{N,0} = Q_{N} + \sum_{j=1}^{k} Q_{j}, k\in [1,N'],
 \end{array} \right. \end{equation}
where $a_k$ are roots of the equations (\ref{xkNxN0-xk0}).
%

Detailed proof of the Lemma 1 can be found in \cite{Lipinski1}.
On Figure 1 it is shown the optimal data transmission graph for the minimum energy problem when
$a\in [a_{k+1},a_{k}]$ and $k\in [1,N']$.
\begin{figure} [!ht] \begin{center}
\includegraphics[width=200pt]{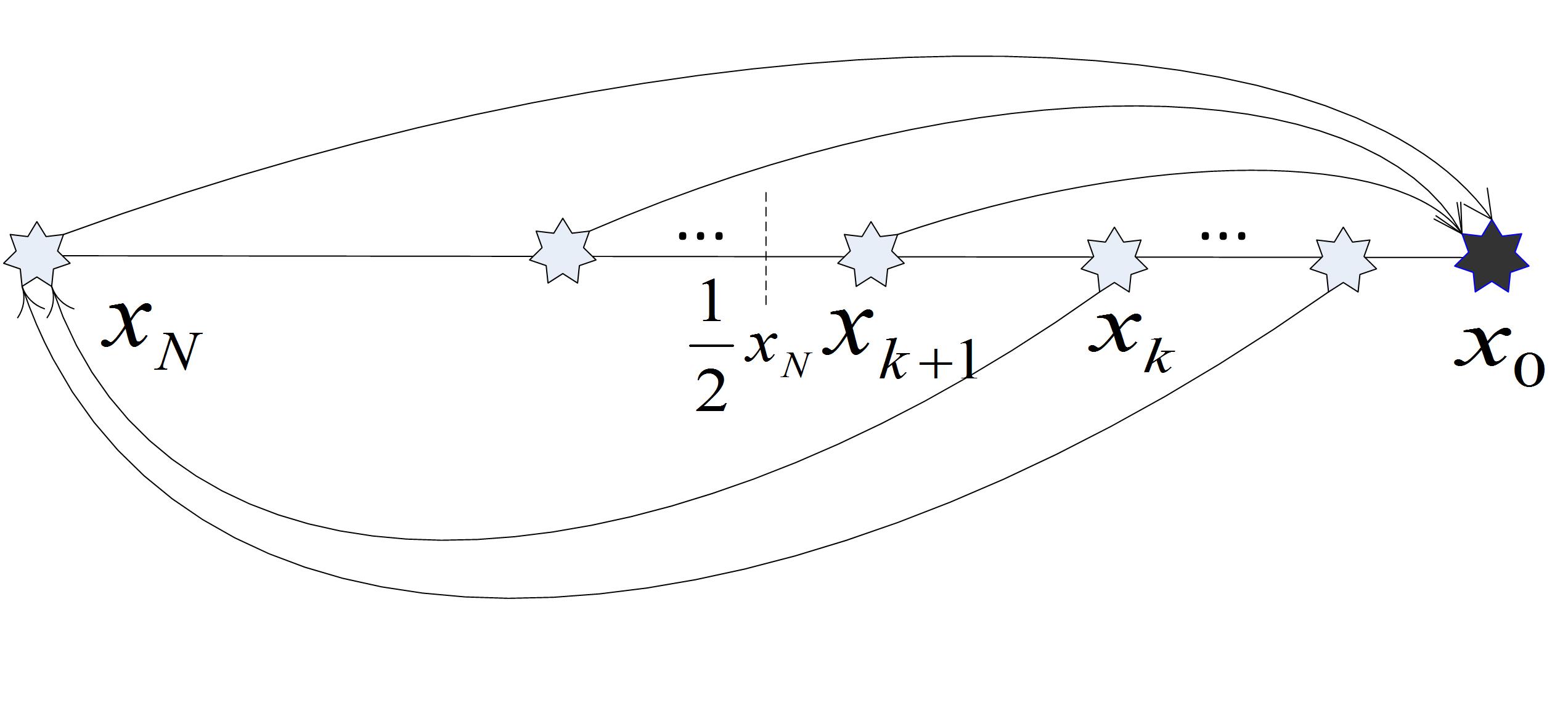}
\caption{Optimal data transmission in $S_N$ for $a\in [a_{k+1},a_{k}]$.}
\label{SolutionTreesMLBP}
\end{center} \end{figure}

The data transmission cost function
\begin{equation} \label{EijLambdaanSN}
E_{i,j}(\bar{\lambda},\bar{\alpha})=\sum_{n}^{} \lambda_n |x_i-x_j|^{\alpha_n}, \;\; \forall_n\; \lambda_n \geq 0,
\end{equation}
which is a sum of monomials (\ref{DataTransCostEnergyMat-SN}) with nonnegative coefficients $\lambda_n$,
for $\forall_n\;\alpha_n\geq 1$ it satisfies (\ref{SuperadditiveDataTransCostEnergMatr})
and
for $\forall_n\; \alpha_n\leq 1$ (\ref{SubadditiveDataTransCostEnergMatr}).
The following lemma defines conditions under which the solution of (\ref{MiniNetEnergyProblem})
for the data transmission cost function (\ref{EijLambdaanSN}) is given by (\ref{SolutionGraph}).

{\bf Lemma 2}.
For the data transmission cost function (\ref{EijLambdaanSN}),
where $\forall_{n}\;\alpha_n\in [a_{k}, a_{k-1}]$, $k\in [1,N']$ and $\lambda_n \geq 0$
the solution of the minimum energy problem is given by the weighted
transmission graph (\ref{SolutionGraph}).

{\it Proof.} We can split the total energy consumed by the network into summands such,
that $E_{T}(q) = \sum_{n} \lambda_n E^n_{T}(q)$, where
$E^n_{T}(q) = \sum_{i\in S_N} E_{i}(q,E^n)$ and $E^{n}_{i,j}=|x_i-x_j|^{\alpha_n}$.
Because for each $\alpha_n\in [a_{k}, a_{k-1}]$ the minimal energy utilized by the sensors is given
by the same transmission graph $T^{(k)}$ from (\ref{SolutionGraph}), then the optimal data transmission
for $E_{i,j}(\bar{\lambda},\bar{\alpha})$ is also given by $T^{(k)}$.$\diamond$
\section{Solution of the problem with the polynomial cost function}
We note that, the objective function $E_{T}(q)$ is linear and continuous
and for this reason its minima lie on the border of the feasible set defined by the second relation in (\ref{MiniNetEnergyProblem}).
From this follows, that the data transmitted by each node cannot be split and it must be sent to a single receiver.
From the Lemma 2 we know, that when the exponents $a, b$ in the data transmission cost function
(\ref{MixedDataTransCostEnMat-SN})
belong to the same interval $[a_{k},a_{k-1}]$,
then the optimal transmission graph for $E_{i,j}= |x_i-x_j|^{a}$ and $E_{i,j}= |x_i-x_j|^{b}$
is the same and it is also optimal for $E_{i,j}= |x_i-x_j|^{a}+\lambda |x_i-x_j|^{b}$.
The problem arises when the solutions of (\ref{MiniNetEnergyProblem})
for two exponents $a$ and $b$ in $E_{i,j}= |x_i-x_j|^{a}$
are given by different data transmission graphs $T_{a}$ and $T_{b}$.
This is because, the optimal transmission graph for the cost function (\ref{MixedDataTransCostEnMat-SN})
cannot be sum of the graphs $T_{a}$ and $T_{b}$, unless
the transmitted data in the graph $T_{a}\cup T_{b}$ are not split.

In this section we consider solutions of (\ref{MiniNetEnergyProblem}) in the one dimensional, regular sensor network $L_N$
for which the nodes are located at the $x_i=i$ points of the half line.
For the $L_N$ network the data transmission cost matrix (\ref{MixedDataTransCostEnMat-SN}) has the form
\begin{equation} \label{MixedDataTransCostEnMat-LN}
	E_{i,j}=|i-j|^a + \lambda |i-j|^b.
\end{equation}
All presented below results are also valid for the non-regular network $S_N$,
but the formulas are cumbersome because of their size and they will not be presented here.

The following lemma describes solution of the the minimum energy problem for the $L_N$ network
when the exponents $a$ and $b$ in (\ref{MixedDataTransCostEnMat-LN}) belong to the neighboring intervals $[a_{k},a_{k-1}]$.

{\bf  Lemma 3}.
For the data transmission cost matrix (\ref{MixedDataTransCostEnMat-LN}),
when
$$
\left\{   \begin{array}{ll}
a\in [1,\infty), & b\in [a_1,1],\\
a\in [a_{k},a_{k-1}], &b\in [a_{k+1}, a_{k}],\;\; k\in [1,N'-1],\\
a\in [a_{N'},a_{N'-1}], &b\in (-\infty, a_{N'}],
 \end{array} \right.$$
the solution of the minimum energy problem is given by
\begin{equation} \label{SolutionMEP}
\left\{   \begin{array}{ll}
T^{(0)}\; {\rm for} \;\lambda \in [0,\lambda_{0} ], &
T^{(1)} \; {\rm for} \;  \lambda \in [\lambda_{0}^{'},\infty),\\
T^{(k)} \; {\rm for} \;  \lambda \in [0,\lambda_{k}], &
T^{(k+1)} \; {\rm for} \;  \lambda \in [\lambda_{k}, \infty),\;\; k\in [1,N'-1],\\
T^{(N')} \; {\rm for} \;  \lambda \in [0,\lambda_{N'}],&
T^{(N'+1)} \; {\rm for} \;  \lambda \in [\lambda_{N'}, \infty),\\
 \end{array} \right.\end{equation}
where
\begin{equation} \label{SolutionMepLambdas}
\left\{   \begin{array}{l}
\lambda_{0} = \frac{2^a - 2}{2 -2^b},\\
\lambda_{0}^{'} = \frac{N^a  - |N-N'-1|^a  - (N'+1)^a}{|N-N'-1|^b + (N'+1)^b - N^b },\\
\lambda_{k} = \frac{(N-k)^a + N^a - k^a}{k^b - (N-k)^b - N^b},\;\; k\in [1,N'],\\
 \end{array}\right.\end{equation}
$a_{0}=1$, $N'=\frac{N-2}{2}$ for $N$ even and $N'=\frac{N-1}{2}$
for $N$ odd,
$T^{(k)}$, $k\in [0,N']$ is a set of data transmission graphs given by (\ref{SolutionGraph}).

{\it Proof.}
If $a\in [a_{k},a_{k-1}]$ and $b\in [a_{k+1},a_{k}]$ in (\ref{MixedDataTransCostEnMat-LN}),
then we know from Lemma 1 that for a sufficiently small $\lambda$  the solution of (\ref{MiniNetEnergyProblem})
is given by the weighted transmission graph
$T^{(k)}=\{ T^{(k)}_{i,N}, T^{(k)}_{i',0} \}_{i=1,i'=k}^{k-1,N}$,
and for a sufficiently large $\lambda$ the solution is given by
$T^{(k+1)}=\{ T^{(k+1)}_{i,N}, T^{(k+1)}_{i',0} \}_{i=1,i'=k+1}^{k,N}$, $k\in[2,N']$.
Because of the linearity and  continuity of the objective function $E_{T}(q)$ its minimum lies on the border of the
feasible set. This means that the data transmitted by each node cannot be split and the optimal transmission graph  for
arbitrary value of the $\lambda$ parameter in  (\ref{MixedDataTransCostEnMat-LN})
cannot be sum of the two graphs $T^{(k)}$ and $T^{(k+1)}$.
To find the optimal transmission for any value of $\lambda$ in (\ref{MixedDataTransCostEnMat-LN}) we order
the transmission graphs in a sequence  such that the cost of data transmission along
$T^{(k)}$ is less or equal the costs along $T^{(k')}$
$$ E_{T}(q^{(k)})\leq  E_{T}(q^{(k')}).$$
The minimal graph $T^{(k')}$ determines the $\lambda_k$ below which the solution of (\ref{MiniNetEnergyProblem}) is given
by weight matrix $q^{(k)}$ of the graph  $T^{(k)}$.
Similarly, the transmission graph $T^{(k'')}$ for which the inequality
$E_{T}(q^{(k+1)})\leq  E_{T}(q^{(k'')})$
is satisfied determines the value of $\lambda_k^{'}$ above which the solution of (\ref{MiniNetEnergyProblem}) is given by $q^{(k+1)}$.
We know from the solution (\ref{SolutionGraph}), (\ref{WeightsForSolutionGraph})
and the inequalities (\ref{SubadditiveDataTransCostEnergMatr}), (\ref{InequalitiesForKandK+1}),
that for the $k$-th node, for any $a\in [a_{k},a_{k-1}]$ and $b\in [a_{k+1},a_{k}]$, $k\in [2,N']$
between $T^{(k)}$  and $T^{(k+1)}$ there is no other optimal data transmission graphs.
For this reason, from the inequality
$$E_{T}(q^{(k)})\leq  E_{T}(q^{(k+1)})$$
we get the values (\ref{SolutionMepLambdas}) of the parameter $\lambda_k$ for which the graphs $T^{(k)}$ and $T^{(k+1)}$ are optimal.

When $a\in [1,\infty)$ and $b\in [a_1, 1]$ in (\ref{MixedDataTransCostEnMat-LN}),
then the solution of (\ref{MiniNetEnergyProblem})  for a sufficiently small $\lambda$
is given by the data transmission graph
$T^{(0)}=\{ T^{(0)}_{i,i-1}\}_{i=1}^{N}$,
and for a sufficiently large $\lambda$ by the transmission graph
$T^{(1)}=\{ T^{(1)}_{i,0}\}_{i=1}^{N}$.
Increasing the parameter  $\lambda$ in (\ref{MixedDataTransCostEnMat-LN}) we pass by set of data transmission
graphs between $T^{(0)}_{}$ and $T^{(1)}_{}$.
The next data transmission graph, which requires more energy then the next hop transmission $T^{(0)}$
is the graph $T^{(0+)}$ in which there is an edge $T^{(0+)}_{N,N-2}$ along the $N$-th node
transmits its $Q_N$ of data to the $(N-2)$ node.
The nodes from $L_N$, $i\in [1,N-1]$ uses the next hop data transmission subgraph with edges $T^{(0)}_{i,i-1}$.
Note that, we cannot select an arbitrary edge $T^{(0+)}_{i,i-2}$ for $i\neq N$.
This is because we want to transmit the minimal amount of data along the edge $T^{(0+)}_{i,i-2}$
and this is satisfied only for the $N$-th node.
The total energy utilized by the network for $T^{(0+)}$ graph is given by the formula
%
$$E_{T}(q^{(0+)}) = E_{T}(q^{(0)}) - E_{N,N-1} - E_{N-1,N-2} + E_{N,N-2}.$$
%
Increasing the value of $\lambda$ in (\ref{MixedDataTransCostEnMat-SN})
we must pass from the transmission graph $T^{(0)}_{}$ to $T^{(0+)}_{}$.
Solving the inequality
$$E_{T}(q^{(0+)})\leq E_{T}(q^{(0)})$$
with respect to the parameter $\lambda$, we get the upper bound  $\lambda \leq \frac{2^a - 2}{2 -2^b}$
given in the lemma.
When we start decrease the value of $\lambda$, above which the data transmission graph
$T^{(1)}_{}$ is optimal, then we pass to the graph $T^{(1+)}$
for which the $N$-th node transmits its $Q_N$ of data along the path which consists of the two edges
$$T^{(1+)}_{N,N-n'-1}, T^{(1+)}_{N-n'-1,0} \in T^{(1+)}_{}.$$
In other words, this is the  transmission path which consists of a one hop of the length
$n'+1$
and the second hop of the length $N-(n'+1)$.
For the $L_N$ network and $k\in [1,N']$,
where $N'$ is the number of nodes in the first part of the network, i.e., $(0,\frac{1}{2}N)$,
$n'=N'$, i.e.,  $N'=\frac{N-2}{2}$ for $N$ even and $N'=\frac{N-1}{2}$ $N$ odd.
The total energy consumed by the network for transmission along the graph $T^{(1+)}$
is given by the formula
%
$$E_{0}(q^{(1+)}_{N}) = E_{0}(q^{(1)}_N) - E_{N,0} + E_{N,N'+1} + E_{N'+1,0}.$$
%
Solving the inequality
$E_{T}(q^{(1+)}_{})\leq E_{T}(q^{(1)}_{}),$
with respect to the parameter $\lambda$, we get the lower bound
$\lambda_{0}^{'}=\frac{N^a  - |N-N'-1|^a  - (N'+1)^a}{|N-N'-1|^b + (N'+1)^b - N^b }$
for which the optimal data transmission graph is $T^{(1)}$.  $\diamond$

The next two lemmas describe the  optimal transmission graphs when the values of $a$ and $b$
in (\ref{MixedDataTransCostEnMat-LN}) does not belong to the
neighboring intervals $[a_{k},a_{k-1}]$.

{\bf Lemma 4}.
Let the exponents of the data transmission cost matrix (\ref{MixedDataTransCostEnMat-LN})
be in the intervals
$a\in [a_{k},a_{k-1}]$ and $b\in [a_{k'},a_{k'-1}]$, $k\in [1,N'-1]$, $k'\in [3,N'+1]$, $k'\geq k+2$,
then the optimal transmission graphs to the minimum energy problem are
$$\left\{   \begin{array}{ll}
T^{(k)}  &  {\rm for} \; \lambda \in (0, \lambda_{k}],\\
T^{(k+i)}  &  {\rm for} \; \lambda \in [\lambda_{k+i-1}, \lambda_{k+i}],  \;\;i\in [1,k'-k-1],\\
T^{(k')}  &  {\rm for} \; \lambda \in [\lambda_{k'-1}, \infty),\\
\end{array} \right.$$
where $\lambda_{k+i-1}$ is the solutions of the inequality
%
$$E_T(q^{(k-1+i)}) \leq E_T(q^{(k+i)}),\;\; \;i \in [1, k'-k].$$
%
{\it Proof.}
We know that, for a sufficiently small $\lambda$ the solution of (\ref{MiniNetEnergyProblem}),
 (\ref{MixedDataTransCostEnMat-LN}), when $a\in [a_{k},a_{k-1}]$ and $b\in [a_{k'},a_{k'-1}]$,
is given by the transmission graph $T^{(k)}$.
From the inequality (\ref{SubadditiveDataTransCostEnergMatr}) and continuity of the objective function $E_{T}$
follows that less optimal, the next to $T^{(k)}$ is the transmission graph $T^{(k+1)}$.
The value of $\lambda_{k}$ above which $T^{(k)}$ is not optimal is determined from the inequality $E_{T}(q^{(k)})\leq E_{T}(q^{(k+1)})$.
Similarly, for a sufficiently large $\lambda$ the solution of (\ref{MiniNetEnergyProblem}) is given
by the graph $T^{(k')}$.
From the inequality (\ref{SubadditiveDataTransCostEnergMatr}) it follows that less optimal,
the closest to $T^{(k')}$, is the transmission graph $T^{(k'-1)}$.
Solving the inequality $E_{T}(q^{(k'-1)})\leq E_{T}(q^{(k')})$ we get the lower bound
of $\lambda_{k'-1}$ for which $T^{(k')}$ is an optimal graph.
By varying the parameter $\lambda$  between $\lambda_{k}$ and $\lambda_{k'-1}$,
when $a\in [a_{k},a_{k-1}]$ and $b\in [a_{k'},a_{k'-1}]$ in  (\ref{MixedDataTransCostEnMat-LN}),
we get the various optimal transmission graphs, different from $T^{(k)}$ and $T^{(k')}$.
From the inequality (\ref{SubadditiveDataTransCostEnergMatr}) and continuity of the objective function $E_{T}$
follows, that the only solutions of (\ref{MiniNetEnergyProblem}) for $\lambda \in [\lambda_{k}, \lambda_{k'}]$
can be transmission graphs $T^{(k+i)}$, $i\in [1,k'-k-1]$.
By solving the set of inequalities $E_T(q^{(k-1+i)}) \leq E_T(q^{(k+i)})$ for $i\in [1,k'-k]$ we get the ordered sequence of
$\lambda_{k+i-1}$.
For any $\lambda$ in the interval $[\lambda_{k+i-1}, \lambda_{k+i}]$ the optimal transmission graph
is $T^{(k+i)}$. $\diamond$

The following lemma  defines the optimal transmission graph when $a\in [1,\infty)$ and $b \leq a_{1}$ in (\ref{MixedDataTransCostEnMat-LN}).

{\bf Lemma 5}.
Let the exponents of (\ref{MixedDataTransCostEnMat-LN}) be in the intervals
$a\in [1,\infty)$ and $b\in [a_{k},a_{k-1}]$, $k\geq 2$, then
the optimal transmission graphs to the minimum energy problem are
$T^{(0)}$ for $\lambda \in [0, \lambda_{0}]$ and
$T^{(k)}$ for $k\geq 2$ $\lambda \in [\lambda_{k}, \infty)$,
where $\lambda_{0}$, $\lambda_{k}$ are given by (\ref{SolutionMepLambdas}).

{\it Proof.} This lemma follows from the Lemma 3.
To get the upper bound $\lambda_0$ of the parameter $\lambda$ for which the transmission graph
$T^{(0)}$ is optimal, we need to solve the inequality $E_{T}(q^{(0+)})\leq E_{T}(q^{(0)})$.
To get the lower bound of the parameter $\lambda$
for which the transmission graphs $T^{(k)}$ are optimal
we have to solve the set of inequalities
$E_{T}(q^{(k-1)})\leq E_{T}(q^{(k)})$, $k\in [2,N'+1]$ which solution $\lambda_{k}$
are given by (\ref{SolutionMepLambdas}). $\diamond$

The optimal transmission graphs of the minimum energy problem when $a\in [a_1, 1]$ and
$b \leq a_{3}$ in (\ref{MixedDataTransCostEnMat-LN})
are for the  parameter $\lambda$.

{\bf Lemma 6}.
Let the exponents of (\ref{MixedDataTransCostEnMat-LN}) be in the intervals
$a\in [a_1, 1]$ and $b\in [a_{k},a_{k-1}]$, $k\geq 3$, then
the optimal transmission graphs to the minimum energy problem are
 $T^{(1)}$ for $\lambda \in [0, \lambda_{0}^{'}]$  and
$T^{(k)}$ for $\lambda \in [\lambda_{k}, \infty)$, $k\in[3,N']$
where $\lambda_{0}^{'}$, $\lambda_{k}$  are given by (\ref{SolutionMepLambdas}). 

{\it Proof.} This lemma follows from the Lemma 3.
To get the upper bound $\lambda_0^{'}$ of the parameter $\lambda$
for which the transmission graph $T^{(1)}$ is optimal
we need to solve the inequality $E_{T}(q^{(1+)})\leq E_{T}(q^{(1)})$.
The lower bound $\lambda_k$ of the of the parameter $\lambda$ for which
the transmission graphs $T^{(k)}$ are optimal can be determined from the
the set of inequalities
$E_{T}(q^{(k-1)})\leq E_{T}(q^{(k)})$, $k\in [3,N'+1]$,
which solution $\lambda_{k}$ are given by (\ref{SolutionMepLambdas}). $\diamond$

\section{Solution of the problem with SINR function}
In previous sections we defined the minimum energy problem in terms
of the data transmission cost matrix $E_{i,j}$ and data flow matrix $q_{i,j}$.
In such formalism there is no information in the model about the data transmission rate,
the sensors operating time and transmission errors caused by the noise and signal interference.
In this section define the minimum energy problem in terms of sensors signal power, data transmission time
and capacities of a transmission channels.
We show, that for the optimal data transmission of the minimum energy problem in the noisy channel
there is no interference of signals.
For omnidirectional antennas, when the signal of the transmitting node is heard in the whole network
this is equivalent to the sequential data transmission.
We prove, that for the point-to-point data transmission utilized by the sensors in the physical layer,
when the transmitter adjust the power of its radio signal to the distance to the receiver,
the solutions of the minimum energy problem coincide with the solutions discussed in the previous sections.

We assume, that the power of the transmitting signal at the receiver
must have some minimal level $P_{0}$.
This requirement means, that the transmitting node must generate the signal with the strength
 \begin{equation} \label{P2P-SignalPower}
          P_{i,j} = P_{0} \;\gamma_{i,j}^{-1},
 \end{equation}
where $\gamma_{i,j}=\gamma(x_i,x_j)$ is the signal gain function between sender and receiver located
at the points  $x_i$ and $x_j$ of the line.
For the transmission model (\ref{P2P-SignalPower}) the energy consumed by the $i$-th sensor is given by the formula
 \begin{equation} \label{NodeEnergySINR}
E_{i}(t)= P_0 \sum_{j\in S_N} \gamma_{i,j}^{-1} t_{i,j},
  \end{equation}
To get non-trivial solution of the minimum energy problem
we must assume that the capacities of the transmission channels are limited,
otherwise the minimum energy of each node is reached for zero transmission time $t_{i,j} =0$.
To define the size of the channel capacity we use the Shannon-Hartley formula modified by
the Signal to Interference plus Noise Ratio (SINR) function, \cite{Gupta, Franceschetti},
%
 \begin{equation} \label{MaxAchievTransRateModel2}
       C(x_i,x_j,U^n_{i,j}) =  \log (1 + s(x_i,x_j,U^n_{i,j})),
 \end{equation}
where
$$s(x_i,x_j,U^n_{i,j})= \frac{P_0}{ \textit{N}_o +  P_0 \sum_{(k,m) \in U^n_{i,j}} \gamma(x_k,x_m)^{-1} \gamma(x_k,x_j)^{}}$$
is the SINR function and $U^n_{i,j}\subset S_N$ is some set of transmitter-receiver pairs which signal of the transmitters interfere
with the signal of the $i$-th node.
For wireless networks in which the nodes use the omnidirectional antennas and the signal is detected
by any node of the network, $U^n_{i,j}$ can be defined as a set of node pairs which transmit data
simultaneously, i.e.,
%
$$
 U_{i,j} = \{ (i',j') \in S_N\times S_N | t^{(s)}_{i,j}=t^{(s)}_{i',j'}, \;t^{(e)}_{i,j}=t^{(e)}_{i',j'} \}.
$$
%
where $t^{(s)}_{i,j}$ and $t^{(e)}_{i,j}$ is the start and the end of transmission time between $i$-th and $j$-th node.
By definition $(i,j) \notin U_{i,j}$.
The amount of data transmitted by the $i$-th node to the $j$-th node during the time  $t_{i,j}$ with
the transmission rate $c_{i,j}$ is given by the formula
 \begin{equation} \label{qij-cij-tij}
            q_{i,j}  = c_{i,j}\; t_{i,j}.
 \end{equation}
We assume, that the transmission rate $c_{i,j}$ satisfies the inequality
%
$ 0 \leq c_{i,j} \leq C(x_i,x_j,U^n_{i,j})$,
%
where $C(x_i,x_j,U^n_{i,j})$ is given by (\ref{MaxAchievTransRateModel2}).
Because in general a set of sensors can transmit data simultaneously, thus we need to modify
the node energy consumption formula (\ref{NodeEnergySINR}) to the form
%
$$ E_{i}(\bar{t})= P_0 \sum_{j, n} \gamma_{i,j}^{-1} \; t^n_{i,j},$$
%
where $\bar{t}=(t^{1}_{}, ... , t^{n}_{}, ... )$ is a tuple of time matrices $t^{n}$ with elements $t^{n}_{i,j}$,
which define the data transmission time between the i-th and j-th nodes in the presence of
transmitters from the set $U^n_{i,j}$.
The objective function of the minimal energy problem with SINR function is given by the formula
%
$$E_{T}(\bar{t}) = \sum_{i\in S_N}^{} E_{i}(\bar{t})= \sum_{i\in S_N}^{}  P_0 \sum_{j, n} \gamma_{i,j}^{-1} \; t^n_{i,j}.$$
%
From the data flow constraints, defined by the second formula in (\ref{MiniNetEnergyProblem}) and (\ref{qij-cij-tij}),
it follows that the minimum energy is consumed by the network
when the transmission rate between two nodes is maximal and equals to the channel capacity, i.e. $c^n_{i,j}=C^n_{i,j}$.
Taking this into account the minimum energy problem with SINR function can be written in the form
 \begin{equation} \label{DataFlowConstraintM2P-SINR} \left\{   \begin{array}{l}
   \min_{\bar{t}}  E_{T}(\bar{t}), \\
   \sum_{i,n} C^n_{i,j} t^n_{i,j} = Q_i + \sum_{j,n} C^n_{j,i} t^n_{j,i},\\
    t^n_{i,j}\geq 0, Q_i>0,
 \end{array} \right.  \end{equation}
where $C^n_{i,j}$  is given by (\ref{MaxAchievTransRateModel2}). 
The results of the following lemma allows us further reduce the problem (\ref{DataFlowConstraintM2P-SINR}).

{\bf Lemma 7}. The optimal data transmission for the minimum energy problem (\ref{DataFlowConstraintM2P-SINR})
is the transmission without interference.

{\it Proof}.
For the fixed amount of data $Q_i$ generated by each sensor, the transmission times $t^n_{i,j}$ in
(\ref{DataFlowConstraintM2P-SINR}) are minimal when coefficients $C^n_{i,j}$ are maximal.
From (\ref{MaxAchievTransRateModel2}) it follows that maximum value of the transmission rate  $C^n_{i,j}$
is achieved when $U^n_{i,j}=\emptyset$, which means that in the network there is no interference of signals. $\diamond$

From the Lemma 7 it follows that to solve the minimum energy problem
it is enough to consider only the constant channel capacities
$   \forall_{i,j}\;\;    C(x_i,x_j) = \log (1 + \frac{P_0}{\textit{N}_o}) = C_0$.
%
The minimum energy problem for noisy channel with the constant channel capacity
 can be defined by the following set of formulas
 \begin{equation} \label{MLProblemM2PNoInterferece}
 \left\{   \begin{array}{l}
  \min_{t}  \sum_{i\in S_N}^{} E_{i}(t), \\
  E_i(t) = P_0 \sum_{j} \gamma_{i,j}^{-1} \; t_{i,j},  \\
  C_0 \sum_{i} t_{i,j} = Q_i + C_0 \sum_{j} t_{j,i},\\
  t_{i,j}\geq 0, Q_i>0.
  \end{array} \right.\end{equation}
To solve the problem (\ref{MLProblemM2PNoInterferece})
for a given signal gain function $\gamma_{i,j}$ we transform (\ref{MLProblemM2PNoInterferece})
to the minimum energy problem defined in (\ref{MiniNetEnergyProblem}).
By identifying the variables
$$\left\{   \begin{array}{l}
q_{i,j}\rightarrow P_0 t_{i,j},\\
E_{i,j} \rightarrow \gamma^{-1}_{i,j},\\
Q_{i} \rightarrow \frac{P_0}{C_0} Q_{i},\\
 \end{array} \right.$$
we get the equivalence of the two problems.
For the signal gain functions $\gamma^{}_{i,j}=|x_i-x_j|^{-a}$,
$\gamma^{}_{i,j}=\frac{1}{|i-j|^{a}+\lambda |i-j|^{b}}$, by means of the above transformation
we can obtain from  (\ref{SolutionGraph}) and (\ref{SolutionMEP}) the
solutions of (\ref{MLProblemM2PNoInterferece}).

\section{Conclusions}
In the paper we solved the minimum energy problem in one dimensional wireless sensor networks for
the data transmission cost function $E(x_i,x_j)=|x_i - x_j|^a$ with any real value of the exponent $a$.
We showed, how to find the solution of the problem when
the data transmission cost function is of the form $E(x_i,x_j)=|x_i - x_j|^a + \lambda |x_i - x_j|^b$
and $a,b\in R$, $\lambda\geq 0$.
There are several intervals for the parameter $\lambda$ for which the optimal transmission graphs are not determined.
For example, when $a\in [1,\infty)$,  $b\in [a_1,1]$ in the interval
$[\lambda_{0}, \lambda_{0}^{'}]$ there are transmission graphs
which lie between $T^{(0)}$ and $T^{(1)}$ and are solutions of (\ref{MiniNetEnergyProblem}).
These graphs can identified by means of the ordering method utilized in the Lemmas 3, 4, 5, 6.
We defined the minimum energy problem in terms of sensors signal power, transmission time and capacities of a transmission channels.
We proved, that for the point-to-point data transmission utilized by the sensors in the physical layer,
when the transmitter adjust the power of its radio signal to the distance to the receiver,
the solutions of the minimum energy problem
written in terms of data transmission cost function $E_{i,j}$
and in terms of sensor signal power coincide.



\begin{thebibliography}{99}
\bibitem{Chang} J.H. Chang, L. Tassiulas, Energy Conserving Routing in Wireless Ad-hoc Networks, Proceedings INFOCOM 2000, pp.22-31.
%
\bibitem{Giridhar} A. Giridhar, P.R. Kumar, Maximizing the functional lifetime of sensor networks,
Proceedings of the 4-th International Symposium on Information Processing in Sensor Networks, Piscataway, NJ, USA, 2005. IEEE Press.
%
\bibitem{Acharya} T. Acharya,  G. Paul, Maximum lifetime broadcast communications in cooperative multihop wireless ad hoc networks: Centralized and distributed approaches, Ad Hoc Networks 11 (2013), pp.1667-1682.
%
\bibitem{Halpern} L. Li, J.Y. Halpern, A Minimum-Energy Path-Preserving Topology-Control Algorithm, IEEE Transaction on Wireless Communications, May, 2004, 910-921.
%
\bibitem{Rodoplu}V. Rodoplu, T.H. Meng, Minimum energy mobile wireless networks, IEEE Journal on selscted areas in communication, vol. 17, no. 8, August 1999.
%
\bibitem{Wu} J. Wu, Handbook on theoretical and algorithmic aspect of sensor, ad hoc wireless, and peer-to-peer networks, Auerbach Publications, 2005.
%
\bibitem{Srinivas} A. Srinivas, E. Modiano, Minimum Energy Disjoint Path Routing in Wireless Ad-hoc Networks,
Proceedings of the 9th Annual International Conference on Mobile Computing and Networking, 2003, p.122-133.
%
\bibitem{Baccelli} F. Baccelli, B. Blaszczyszyn, Stochastic Geometry and Wireless Networks, vol. 1, 2, Now Publishers Inc, 2009.
%
\bibitem{Garey} M. Garey, D. Johnson, Computers and Intractability: a guide to theory of NP-Completeness, Freeman, San Francisco, USA, 1979.
%
\bibitem{Steele} M.J. Steele, Probability theory and combinatorial optimization, SIAM, Philadelphia, 1997.
%
\bibitem{Lipinski1}
Z. Lipi{\'n}ski, On classification of data transmission strategies in one dimensional wireless ad-hoc networks
with polynomial cost function, Monographs Of System Dependability, DepCoS-RELCOMEX 2012, Poland, pp. 85-104.
%
\bibitem{Gupta} P. Gupta, P. R. Kumar, The Capacity of Wireless Networks, IEEE Transactions on Information Theory, 46(2), pp.388-404, 2000.
%
\bibitem{Franceschetti} M. Franceschetti, R. Meester, Random Networks for Communication, Cambridge University Press, 2007.
%
\end{thebibliography}
\end{document}